# New Perspective on the Optical Theorem of Classical Electrodynamics

Masud Mansuripur

College of Optical Sciences, The University of Arizona, Tucson, Arizona 85721

masud@optics.arizona.edu

**Abstract**. A general proof of the optical theorem (also known as the optical cross-section theorem) is presented that reveals the intimate connection between the forward scattering amplitude and the absorption-plus-scattering of the incident wave within the scatterer. The oscillating electric charges and currents as well as the electric and magnetic dipoles of the scatterer, driven by an incident plane-wave, extract energy from the incident beam at a certain rate. The same oscillators radiate electro-magnetic energy into the far field, thus giving rise to well-defined scattering amplitudes along various directions. The essence of the proof presented here is that the extinction cross-section of an object can be related to its forward scattering amplitude using the induced oscillations within the object but *without* an actual knowledge of the mathematical form assumed by these oscillations.

**1. Introduction**. The optical theorem relates the extinction cross-section of an arbitrary object placed in the path of a monochromatic plane-wave to its forward scattering amplitude, namely, the scattered light amplitude measured in the far field along the propagation direction (and at the frequency of) the incident plane-wave [1,2]. The scatterer could have arbitrary geometric shape and electromagnetic profile – for example, it could be inhomogeneous, anisotropic, dispersive, absorptive, nonlinear, magnetic, etc., with arbitrarily complex constitutive relations governing its electromagnetic response. The incident beam feeds energy to the scatterer at a certain rate, which energy may be converted to other forms (e.g., thermal, chemical), or re-radiated by the scatterer at various frequencies and in different directions. The extinction cross-section of the scatterer is a measure of its rate of uptake of electromagnetic energy from the incident plane-wave, irrespective of whether that energy is permanently absorbed or subsequently re-radiated into the surrounding space. The beauty of the optical theorem is in the simplicity and generality of the relation between the extinction cross-section and the forward scattering amplitude, the complexity of the interaction between the scatterer and the incident wave notwithstanding.

The first inklings of the optical theorem appear in the work of J.W. Strutt (Lord Rayleigh) as far back as 1871 [3]. The theorem is stated explicitly in the famous 1908 paper by G. Mie on the scattering of light by spherical particles [4]. In 1943, W. Heisenberg invented the *S* matrix as a quantum-theoretical tool for description of the scattering of particles, apparently unaware of prior work in this area by J.A. Wheeler and also by N. Bohr, R.E. Peierls, and G. Placzek. Heisenberg then proceeded to prove the unitarity of the *S* matrix and, as an observable consequence of this rather abstract property, derived the *generalized optical theorem* [5]. Back in electromagnetic theory, H.C. van de Hulst rediscovered the optical theorem in 1949 [6], unaware that it was already well known both in optics and in quantum scattering theory. The optics treatise by M. Born and E. Wolf [2] nevertheless credits van de Hulst with the first derivation of the theorem in the domain of classical optics. For a more detailed discussion of the history and significance of the theorem, see R.G. Newton [7], who also reproduces van de Hulst's scalar derivation as a "very nice and intuitive" exposition of the theorem "from a physical point of view," albeit one that lacks the generality and accuracy of the full vector derivation.

The objective of the present paper is an exact derivation of the optical theorem in the electromagnetic domain, starting with Maxwell's equations and without undue reliance on mathematical theorems in the process of derivation. In Sec. 2, introducing the electric charge and current distributions as well as the distributions of electric and magnetic dipoles induced within the scatterer, we relate their oscillations (at the frequency of the incident wave) to the rate of uptake of energy by the scatterer from the incident *E*- and *H*-fields. The same oscillations also give rise to the scattered light, whose far field amplitude along the propagation direction of the incident wave will be derived in Sec. 3. A comparison of the extinction cross-section derived in Sec. 2 with the forward scattering amplitude derived in Sec. 3 reveals the intimate connection between the two, despite the fact that the actual form of the internal oscillations of the scatterer remains unknown. Section 4 describes a simple application of the optical theorem in the case of an opaque, flat object illuminated at normal incidence. Some final thoughts and concluding remarks are relegated to Sec. 5.

**2. Absorption-plus-scattering cross-section**. With reference to Fig.1, consider a plane, monochromatic wave of frequency $\omega_o$, propagating in free space along the *z*-axis. The electromagnetic field amplitudes are given by

$$\boldsymbol{E}(\boldsymbol{r},t) = \mathrm{Re}\{\tilde{\boldsymbol{E}}_o \exp[\mathrm{i}(k_o z - \omega_o t)]\}, \tag{1a}$$

$$\boldsymbol{H}(\boldsymbol{r},t) = \mathrm{Re}\{\hat{\boldsymbol{z}} \times (\tilde{\boldsymbol{E}}_o / Z_o)\exp[\mathrm{i}(k_o z - \omega_o t)]\}. \tag{1b}$$

Here $\tilde{\boldsymbol{E}}_o = \tilde{E}_{xo}\hat{\boldsymbol{x}} + \tilde{E}_{yo}\hat{\boldsymbol{y}} = (E'_{xo} + \mathrm{i}E''_{xo})\hat{\boldsymbol{x}} + (E'_{yo} + \mathrm{i}E''_{yo})\hat{\boldsymbol{y}}$ is the complex *E*-field amplitude, $\omega_o$ is the angular frequency, $k_o = \omega_o/c$ is the magnitude of the *k*-vector, $c = 1/\sqrt{\mu_o\varepsilon_o}$ is the speed of light in vacuum, and $Z_o = \sqrt{\mu_o/\varepsilon_o}$ is the impedance of free space; $\mu_o$ and $\varepsilon_o$ are the usual permeability and permittivity of free space in the MKSA system of units adopted here. In general, a tilde (~) under a symbol indicates that the symbol represents a complex-valued entity. Note that the complex *E*-field amplitude $\tilde{\boldsymbol{E}}_o$ is confined to the *xy*-plane, but is otherwise arbitrary; in particular, the presence of the real and imaginary vectors $\boldsymbol{E}'_o = E'_{xo}\hat{\boldsymbol{x}} + E'_{yo}\hat{\boldsymbol{y}}$ and $\boldsymbol{E}''_o = E''_{xo}\hat{\boldsymbol{x}} + E''_{yo}\hat{\boldsymbol{y}}$ ensures that all possible states of polarization are realized. The magnetic field vector $\tilde{\boldsymbol{H}}_o = \hat{\boldsymbol{z}} \times \tilde{\boldsymbol{E}}_o / Z_o$ is also confined to the *xy*-plane, its real and imaginary parts being derived from the corresponding components of $\tilde{\boldsymbol{E}}_o$ by cross-multiplication with $\hat{\boldsymbol{z}}$ and division by $Z_o$.

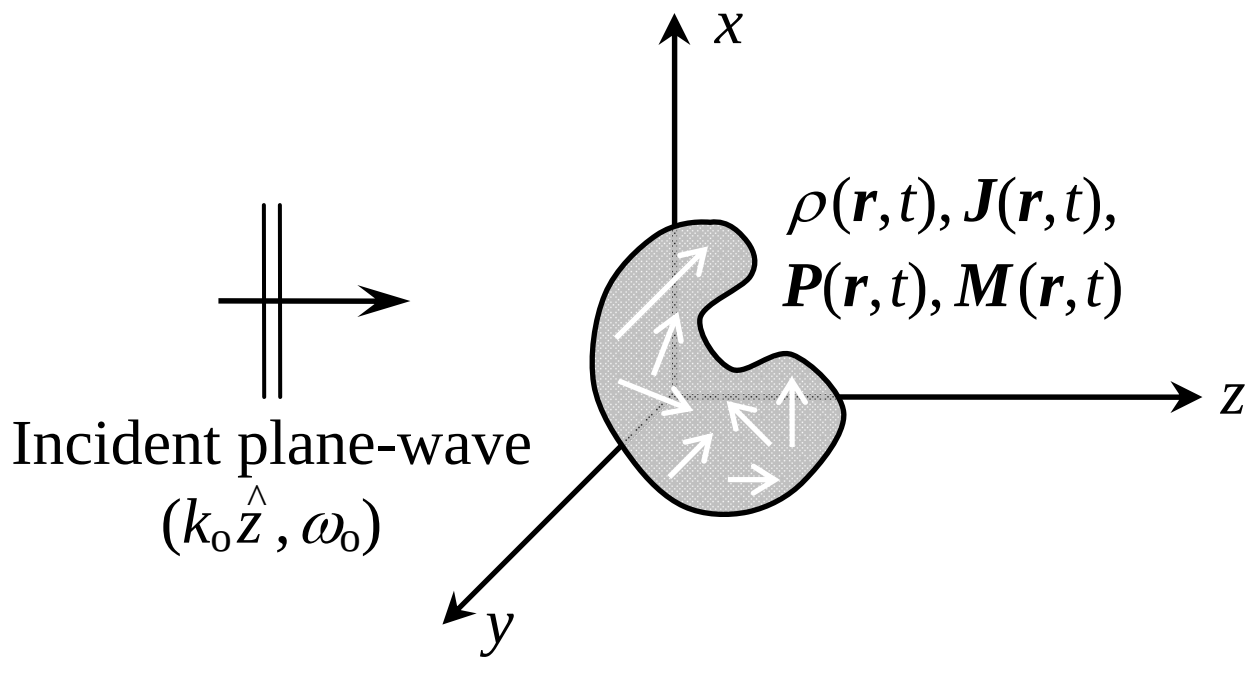

**Fig. 1**. A plane, monochromatic, electromagnetic wave propagating along the *z*-axis excites electric charge, electric current, polarization, and magnetization within a material object. The oscillations thus induced within the material body give rise to the scattered electromagnetic field permeating the space inside as well as outside the object.

The plane-wave described by Eq. (1) illuminates a material body placed at the origin of the coordinate system. The material object responds by developing throughout its volume a polarization $\boldsymbol{P}(\boldsymbol{r},t)$, a magnetization $\boldsymbol{M}(\boldsymbol{r},t)$, a conduction electron current-density $\boldsymbol{J}(\boldsymbol{r},t)$, and a conduction electron charge-density $\rho(\boldsymbol{r},t)$. Interaction between the incident plane-wave and the induced polarization, magnetization, and current-density causes a certain amount of optical energy to be taken away from the incident wave and delivered to the material object. According to Poynting's theorem [1,8,9], the rate of exchange of energy density, $\partial\mathcal{E}(\boldsymbol{r},t)/\partial t$, is given by

$$\partial\mathcal{E}(\boldsymbol{r},t)/\partial t = \boldsymbol{E}(\boldsymbol{r},t)\cdot\boldsymbol{J}(\boldsymbol{r},t) + \boldsymbol{E}(\boldsymbol{r},t)\cdot\partial\boldsymbol{P}(\boldsymbol{r},t)/\partial t + \boldsymbol{H}(\boldsymbol{r},t)\cdot\partial\boldsymbol{M}(\boldsymbol{r},t)/\partial t. \qquad (2)$$

Note that the *E*- and *H*-fields appearing in Eq. (2) are the fields of the incident plane-wave given by Eq. (1). The electric and magnetic fields produced by the oscillations of the material body (the so-called self fields) also exchange energy with the polarization, magnetization, and current-density, but these exchanges do *not* contribute to the extinction cross-section of the object; see Sec. 5 for a detailed discussion of this point.

When placed in Eq. (2) and integrated over a long time interval, the contributions to $\partial\mathcal{E}(\boldsymbol{r},t)/\partial t$ of the material oscillations with frequencies other than $\omega_o$ will average out to zero. (Such frequencies do not arise in a linear medium, but a nonlinear scatterer could exhibit oscillations at $\omega \neq \omega_o$.) Only those oscillations having the same frequency as that of the incident plane-wave will contribute to the time-averaged energy exchange rate, $<\partial\mathcal{E}(\boldsymbol{r},t)/\partial t>$. The relevant material oscillations may therefore be expressed as the real parts of $\underset{\sim}{\boldsymbol{P}}(\boldsymbol{r})\exp(-\mathrm{i}\omega_o t)$, $\underset{\sim}{\boldsymbol{M}}(\boldsymbol{r})\exp(-\mathrm{i}\omega_o t)$, $\underset{\sim}{\boldsymbol{J}}(\boldsymbol{r})\exp(-\mathrm{i}\omega_o t)$, and $\underset{\sim}{\rho}(\boldsymbol{r})\exp(-\mathrm{i}\omega_o t)$. Substitution into Eq. (2) and time-averaging now yields

$$<\partial\mathcal{E}(\boldsymbol{r},t)/\partial t> = \tfrac{1}{2}\mathrm{Re}\left\{[\underset{\sim}{\boldsymbol{E}}_o^*\cdot\underset{\sim}{\boldsymbol{J}}(\boldsymbol{r}) - \mathrm{i}\omega_o\underset{\sim}{\boldsymbol{E}}_o^*\cdot\underset{\sim}{\boldsymbol{P}}(\boldsymbol{r}) - \mathrm{i}\omega_o\underset{\sim}{\boldsymbol{H}}_o^*\cdot\underset{\sim}{\boldsymbol{M}}(\boldsymbol{r})]\exp(-\mathrm{i}k_o z)\right\}\cdot \qquad (3)$$

When integrated over the volume of the material body – or, equivalently, over the entire space – Eq. (3) yields the time-averaged rate of absorption-plus-scattering of the incident beam by the material object. Considering that the energy flux of the incident plane-wave is given by the corresponding Poynting vector $\boldsymbol{S}(\boldsymbol{r},t) = \boldsymbol{E}(\boldsymbol{r},t)\times\boldsymbol{H}(\boldsymbol{r},t)$, we will have

$$<\boldsymbol{S}(\boldsymbol{r},t)> = \frac{1}{2Z_o}\mathrm{Re}(\underset{\sim}{\boldsymbol{E}}_o\cdot\underset{\sim}{\boldsymbol{E}}_o^*)\,\hat{\boldsymbol{z}} = \frac{1}{2Z_o}(\boldsymbol{E}_o'\cdot\boldsymbol{E}_o' + \boldsymbol{E}_o''\cdot\boldsymbol{E}_o'')\,\hat{\boldsymbol{z}}. \qquad (4)$$

The absorption-plus-scattering (i.e., extinction) cross-section $\mathcal{A}$ of the object is thus given by

$$\mathcal{A} = \frac{\mathrm{Re}\left\{\underset{\sim}{\boldsymbol{E}}_o^*\cdot\int_{-\infty}^{\infty}\{Z_o[\underset{\sim}{\boldsymbol{J}}(\boldsymbol{r}) - \mathrm{i}\omega_o\underset{\sim}{\boldsymbol{P}}(\boldsymbol{r})] - \mathrm{i}\omega_o\underset{\sim}{\boldsymbol{M}}(\boldsymbol{r})\times\hat{\boldsymbol{z}}\}\exp(-\mathrm{i}k_o z)\mathrm{d}\boldsymbol{r}\right\}}{|\boldsymbol{E}_o'|^2 + |\boldsymbol{E}_o''|^2}\cdot \qquad (5)$$

So far as the absorption and scattering processes are concerned, the above extinction cross-section $\mathcal{A}$ is the effective area (projected onto the *xy*-plane) that the object presents to the incident plane-wave. A flat, perfectly absorbing obstacle of area $\mathcal{A}$ placed in the *xy*-plane, would *absorb* energy at the same rate from the incident beam as the aforementioned material body takes away via both absorption and scattering processes.

**3. Forward-scattered electromagnetic field**. The distribution of total charge-density (free + bound) within and on the surface(s) of the object is given by $\rho(\boldsymbol{r},t)-\boldsymbol{\nabla}\cdot\boldsymbol{P}(\boldsymbol{r},t)$. The charges oscillating at frequency $\omega_0$ produce the following scalar potential $\psi(r,t)$ in the Lorenz gauge:

$$\psi(\boldsymbol{r},t)=\mathrm{Re}[\underset{\sim}{\psi}(\boldsymbol{r})\exp(-\mathrm{i}\omega_\mathrm{o}t)]=\mathrm{Re}\int_{-\infty}^{\infty}\frac{\underset{\sim}{\rho}(\boldsymbol{r}')-\boldsymbol{\nabla}\cdot\underset{\sim}{\boldsymbol{P}}(\boldsymbol{r}')}{4\pi\varepsilon_\mathrm{o}|\boldsymbol{r}-\boldsymbol{r}'|}\exp(\mathrm{i}\omega_\mathrm{o}|\boldsymbol{r}-\boldsymbol{r}'|/c)\exp(-\mathrm{i}\omega_\mathrm{o}t)\,\mathrm{d}\boldsymbol{r}'. \quad (6)$$

Similarly, the object's total current-density distribution, $\boldsymbol{J}(\boldsymbol{r},t)+\partial\boldsymbol{P}(\boldsymbol{r},t)/\partial t+\mu_\mathrm{o}^{-1}\boldsymbol{\nabla}\times\boldsymbol{M}(\boldsymbol{r},t)$, gives rise to the following vector potential, also in the Lorenz gauge, at the frequency $\omega_0$ [1,2,8]:

$$\boldsymbol{A}(\boldsymbol{r},t)=\mathrm{Re}[\underset{\sim}{\boldsymbol{A}}(\boldsymbol{r})\exp(-\mathrm{i}\omega_\mathrm{o}t)]$$
$$=\mathrm{Re}\int_{-\infty}^{\infty}\frac{\mu_\mathrm{o}[\underset{\sim}{\boldsymbol{J}}(\boldsymbol{r}')-\mathrm{i}\omega_\mathrm{o}\underset{\sim}{\boldsymbol{P}}(\boldsymbol{r}')]+\boldsymbol{\nabla}\times\underset{\sim}{\boldsymbol{M}}(\boldsymbol{r}')}{4\pi|\boldsymbol{r}-\boldsymbol{r}'|}\exp(\mathrm{i}\omega_\mathrm{o}|\boldsymbol{r}-\boldsymbol{r}'|/c)\exp(-\mathrm{i}\omega_\mathrm{o}t)\,\mathrm{d}\boldsymbol{r}'. \quad (7)$$

The scattered $E$-field may now be obtained from $\boldsymbol{E}_\mathrm{s}(\boldsymbol{r},t)=-\boldsymbol{\nabla}\psi(\boldsymbol{r},t)-\partial\boldsymbol{A}(\boldsymbol{r},t)/\partial t.$ Writing $\boldsymbol{E}_\mathrm{s}(\boldsymbol{r},t)=\mathrm{Re}[\underset{\sim}{\boldsymbol{E}}_\mathrm{s}(\boldsymbol{r})\exp(-\mathrm{i}\omega_\mathrm{o}t)],$ the complex $E$-field amplitude at the frequency $\omega_0$ will be

$$\underset{\sim}{\boldsymbol{E}}_\mathrm{s}(\boldsymbol{r})=\int_{-\infty}^{\infty}\left\{\frac{[\underset{\sim}{\rho}(\boldsymbol{r}')-\boldsymbol{\nabla}\cdot\underset{\sim}{\boldsymbol{P}}(\boldsymbol{r}')](1-\mathrm{i}k_\mathrm{o}|\boldsymbol{r}-\boldsymbol{r}'|)(\boldsymbol{r}-\boldsymbol{r}')}{4\pi\varepsilon_\mathrm{o}|\boldsymbol{r}-\boldsymbol{r}'|^3}+\mathrm{i}\omega_\mathrm{o}\frac{\mu_\mathrm{o}[\underset{\sim}{\boldsymbol{J}}(\boldsymbol{r}')-\mathrm{i}\omega_\mathrm{o}\underset{\sim}{\boldsymbol{P}}(\boldsymbol{r}')]+\boldsymbol{\nabla}\times\underset{\sim}{\boldsymbol{M}}(\boldsymbol{r}')}{4\pi|\boldsymbol{r}-\boldsymbol{r}'|}\right\}$$
$$\times\exp(\mathrm{i}k_\mathrm{o}|\boldsymbol{r}-\boldsymbol{r}'|)\,\mathrm{d}\boldsymbol{r}'. \quad (8)$$

The term containing $\underset{\sim}{\boldsymbol{M}}(\boldsymbol{r}')$ in the above integral requires further algebraic manipulations. Using the vector identity $\phi\boldsymbol{\nabla}\times\boldsymbol{V}=\boldsymbol{\nabla}\times(\phi\boldsymbol{V})-\boldsymbol{\nabla}\phi\times\boldsymbol{V}$, we rewrite this term as follows:

$$\int_{-\infty}^{\infty}[\exp(\mathrm{i}k_\mathrm{o}|\boldsymbol{r}-\boldsymbol{r}'|)/|\boldsymbol{r}-\boldsymbol{r}'|]\boldsymbol{\nabla}\times\underset{\sim}{\boldsymbol{M}}(\boldsymbol{r}')\,\mathrm{d}\boldsymbol{r}'=\int_{-\infty}^{\infty}\boldsymbol{\nabla}\times[\exp(\mathrm{i}k_\mathrm{o}|\boldsymbol{r}-\boldsymbol{r}'|)\underset{\sim}{\boldsymbol{M}}(\boldsymbol{r}')/|\boldsymbol{r}-\boldsymbol{r}'|]\,\mathrm{d}\boldsymbol{r}'$$
$$-\int_{-\infty}^{\infty}[(1-\mathrm{i}k_\mathrm{o}|\boldsymbol{r}-\boldsymbol{r}'|)\exp(\mathrm{i}k_\mathrm{o}|\boldsymbol{r}-\boldsymbol{r}'|)/|\boldsymbol{r}-\boldsymbol{r}'|^3](\boldsymbol{r}-\boldsymbol{r}')\times\underset{\sim}{\boldsymbol{M}}(\boldsymbol{r}')\,\mathrm{d}\boldsymbol{r}'. \quad (9)$$

Later on, we will need to dot-multiply $\underset{\sim}{\boldsymbol{E}}_\mathrm{s}(\boldsymbol{r})$ of Eq.(8) into the amplitude $\underset{\sim}{\boldsymbol{E}}_\mathrm{o}^*$ of the incident wave, in order to arrive at an expression similar to that in Eq.(5). When this operation is performed on Eq.(9), with the aid of the vector identity $\boldsymbol{U}\cdot(\boldsymbol{\nabla}\times\boldsymbol{V})=\boldsymbol{V}\cdot(\boldsymbol{\nabla}\times\boldsymbol{U})-\boldsymbol{\nabla}\cdot(\boldsymbol{U}\times\boldsymbol{V})$ and Gauss's theorem, namely, $\int_{\text{volume}}\boldsymbol{\nabla}\cdot\boldsymbol{W}(\boldsymbol{r}')\,\mathrm{d}\boldsymbol{r}'=\int_{\text{surface}}\boldsymbol{W}(\boldsymbol{r}')\cdot\mathrm{d}\boldsymbol{s}$, the first integral on the right-hand-side of Eq.(9) will vanish (because the magnetization $\boldsymbol{M}$ outside the material body is zero). The product $\underset{\sim}{\boldsymbol{E}}_\mathrm{o}^*\cdot\underset{\sim}{\boldsymbol{E}}_\mathrm{s}(\boldsymbol{r})$ thus becomes

$$\underset{\sim}{\boldsymbol{E}}_\mathrm{o}^*\cdot\underset{\sim}{\boldsymbol{E}}_\mathrm{s}(\boldsymbol{r})=(4\pi)^{-1}\underset{\sim}{\boldsymbol{E}}_\mathrm{o}^*\cdot\int_{-\infty}^{\infty}\left\{\frac{[\underset{\sim}{\rho}(\boldsymbol{r}')-\boldsymbol{\nabla}\cdot\underset{\sim}{\boldsymbol{P}}(\boldsymbol{r}')](1-\mathrm{i}k_\mathrm{o}|\boldsymbol{r}-\boldsymbol{r}'|)(\boldsymbol{r}-\boldsymbol{r}')}{\varepsilon_\mathrm{o}|\boldsymbol{r}-\boldsymbol{r}'|^3}+\mathrm{i}\omega_\mathrm{o}\frac{\mu_\mathrm{o}[\underset{\sim}{\boldsymbol{J}}(\boldsymbol{r}')-\mathrm{i}\omega_\mathrm{o}\underset{\sim}{\boldsymbol{P}}(\boldsymbol{r}')]}{|\boldsymbol{r}-\boldsymbol{r}'|}\right.$$
$$\left.-\,\mathrm{i}\omega_\mathrm{o}\frac{(1-\mathrm{i}k_\mathrm{o}|\boldsymbol{r}-\boldsymbol{r}'|)(\boldsymbol{r}-\boldsymbol{r}')\times\underset{\sim}{\boldsymbol{M}}(\boldsymbol{r}')}{|\boldsymbol{r}-\boldsymbol{r}'|^3}\right\}\exp(\mathrm{i}k_\mathrm{o}|\boldsymbol{r}-\boldsymbol{r}'|)\,\mathrm{d}\boldsymbol{r}'. \quad (10)$$

All the results obtained up to this point are exact. We now invoke the far field approximation by considering the scattered field at a distant point $\boldsymbol{r}=\zeta_\mathrm{o}\hat{\boldsymbol{z}}$ along the $z$-axis (i.e., along the propagation direction of the incident wave). Since $\zeta_o$ is much greater than the dimensions of the object under consideration, we approximate $\boldsymbol{r}-\boldsymbol{r}'$ in Eq.(10) with $\zeta_o\hat{\boldsymbol{z}}$ everywhere except in the exponential term, where we use

$$|\boldsymbol{r}-\boldsymbol{r}'| = [x'^2 + y'^2 + (\zeta_o - z')^2]^{1/2} \approx (\zeta_o - z') + \tfrac{1}{2}(x'^2 + y'^2)/(\zeta_o - z') \approx \zeta_o - z'. \qquad (11)$$

The term $\underset{\sim}{\boldsymbol{E}}_o^* \cdot (\boldsymbol{r}-\boldsymbol{r}') \approx \zeta_o \underset{\sim}{\boldsymbol{E}}_o^* \cdot \hat{\boldsymbol{z}}$ appearing in the first integrand of Eq.(10) now vanishes because $\boldsymbol{E}_o$ has no component along the $z$-axis. Equation (10) may thus be written as follows:

$$\underset{\sim}{\boldsymbol{E}}_o^* \cdot \underset{\sim}{\boldsymbol{E}}_s(\zeta_o\hat{\boldsymbol{z}}) \approx (4\pi)^{-1} \underset{\sim}{\boldsymbol{E}}_o^* \cdot \int_{-\infty}^{\infty} \left\{ \frac{\mathrm{i}\omega_o\mu_o[\underset{\sim}{\boldsymbol{J}}(\boldsymbol{r}') - \mathrm{i}\omega_o \underset{\sim}{\boldsymbol{P}}(\boldsymbol{r}')]}{\zeta_o} - \frac{\mathrm{i}\omega_o(1-\mathrm{i}k_o\zeta_o)\hat{\boldsymbol{z}} \times \underset{\sim}{\boldsymbol{M}}(\boldsymbol{r}')}{\zeta_o^2} \right\} \exp[\mathrm{i}k_o(\zeta_o - z')]\mathrm{d}\boldsymbol{r}'. \qquad (12)$$

Neglecting the term that decays as $1/\zeta_o^2$ with an increasing distance $\zeta_o$ to the far-field observation point, the above equation becomes

$$\underset{\sim}{\boldsymbol{E}}_o^* \cdot \underset{\sim}{\boldsymbol{E}}_s(\zeta_o\hat{\boldsymbol{z}}) \approx \frac{\mathrm{i}k_o \exp(\mathrm{i}k_o\zeta_o)}{4\pi\zeta_o} \underset{\sim}{\boldsymbol{E}}_o^* \cdot \int_{-\infty}^{\infty} \{ Z_o[\underset{\sim}{\boldsymbol{J}}(\boldsymbol{r}') - \mathrm{i}\omega_o \underset{\sim}{\boldsymbol{P}}(\boldsymbol{r}')] - \mathrm{i}\omega_o \underset{\sim}{\boldsymbol{M}}(\boldsymbol{r}') \times \hat{\boldsymbol{z}} \} \exp(-\mathrm{i}k_o z')\mathrm{d}\boldsymbol{r}'. \qquad (13)$$

The (unnormalized) vectorial scattering amplitude $\boldsymbol{F}(\boldsymbol{k},\boldsymbol{k}_o)$ is defined as

$$\underset{\sim}{\boldsymbol{E}}_s(\zeta_o\hat{\boldsymbol{z}}) = \zeta_o^{-1}\exp(\mathrm{i}k_o\zeta_o)\underset{\sim}{\boldsymbol{F}}(\boldsymbol{k}=\boldsymbol{k}_o, \boldsymbol{k}_o), \qquad (14)$$

where $\boldsymbol{k}$ is the wave-vector in the direction of observation, and $\boldsymbol{k}_o = k_o\hat{\boldsymbol{z}}$ is the incident wave-vector. Comparing Eq.(5) with Eq.(13), we find

$$\mathcal{A} = \frac{4\pi\,\mathrm{Im}\{\underset{\sim}{\boldsymbol{E}}_o^* \cdot \underset{\sim}{\boldsymbol{F}}(\boldsymbol{k}=k_o\hat{\boldsymbol{z}}, k_o\hat{\boldsymbol{z}})\}}{k_o\left(|\boldsymbol{E}_o'|^2 + |\boldsymbol{E}_o''|^2\right)}. \qquad (15)$$

Equation (15) is the general statement of the optical theorem, which relates the extinction (i.e., absorption-plus-scattering) cross-section $\mathcal{A}$ of an arbitrary object to its normalized scattering amplitude $\boldsymbol{F}(\boldsymbol{k}=\boldsymbol{k}_o,\boldsymbol{k}_o)$ along the direction of incidence. Note that the field at the observation point is a superposition of the incident plane-wave and the field scattered by the object. To find the scattering amplitude $\boldsymbol{F}(\boldsymbol{k},\boldsymbol{k}_o)$, one must first subtract the contribution of the incident plane-wave, namely, $\underset{\sim}{\boldsymbol{E}}_o\exp(\mathrm{i}k_o\zeta_o)$, from the complex vectorial $E$-field at $\boldsymbol{r}=\zeta_o\hat{\boldsymbol{z}}$, before proceeding to normalize by $\zeta_o^{-1}\exp(\mathrm{i}k_o\zeta_o)$ in accordance with Eq.(14).

**4. Scalar diffraction from a flat, opaque object illuminated at normal incidence**. As an example of the application of Eq.(15), consider an opaque, flat object of area $A_o$ placed perpendicular to the propagation direction of the incident beam within the $xy$-plane. The geometry of the system is the same as that depicted in Fig.1, except that the object from which the incident beam is scattered is now flat and perfectly opaque. Invoking Babinet's principle of the scalar diffraction theory [1,2], we observe that the scattered (or diffracted) field is produced by a uniform $E$-field of amplitude $-\mathrm{Re}[\underset{\sim}{\boldsymbol{E}}_o\exp(-\mathrm{i}\omega_o t)]$ confined to the area $A_o$ occupied by the flat object within the $xy$-plane. The 2-dimensional Fourier transform of the above field profile, evaluated at the spatial frequency $(\sigma_x, \sigma_y) = (x/\zeta_o, y/\zeta_o) = (0,0)$ — corresponding to the on-axis observation point $\boldsymbol{r} = \zeta_o \hat{\boldsymbol{z}}$ in the far field — will then be proportional to the area $A_o$ of the object. Consequently, the Fraunhofer diffraction formula [2,10] yields the scattered field amplitude as follows:

$$\tilde{\boldsymbol{E}}_s(\zeta_o \hat{\boldsymbol{z}}) = -(\mathrm{i}\lambda_o \zeta_o)^{-1} \exp(\mathrm{i}k_o \zeta_o) A_o \tilde{\boldsymbol{E}}_o. \quad (16)$$

Substitution into Eq. (15) reveals that $\mathcal{A} = 2A_o$, that is, the extinction cross-section of the flat, opaque object is twice as large as its geometric area. This is a surprising but well-known result of the classical theory of scattering, stating that precisely one-half of the cross-section $\mathcal{A}$ of an opaque flat object is due to absorption, while the remaining half is due to scattering (i.e., diffraction) from the boundaries, irrespective of the shape and orientation of the object within the *xy*-plane [2].

**5. Concluding remarks**. We have proven the optical theorem of classical electrodynamics using a rigorous yet physically tractable technique. Instead of relying on the integral of the Poynting vector over a closed surface surrounding the scatterer, which is the usual approach to evaluating the extinction cross-section [1,2], we have "peered" into the scatterer and calculated the rate of energy uptake by its internal oscillators. Interestingly, we find that the self-field of the scatterer makes no contribution whatsoever to the extinction cross-section; in other words, the interaction between the incident wave and the oscillators is all that is needed to account for the overall absorption-plus-scattering of the incident beam by the object. This is *not* to say that the self-field does not exchange energy with the oscillators; rather, where the self-field extracts energy from these oscillators, that energy is transformed into the scattered field energy, and where the self-field delivers energy to the oscillators, that is energy that is already accounted for, since it has been supplied by the incident beam in the first place. The recognition that the self-field does *not* contribute to the extinction cross-section is the crucial physical insight of the present paper.

The second half of the paper is concerned with the calculation of the forward scattering amplitude. The traditional methods use Green's theorem along with certain mathematical lemmas to relate the far field amplitude to the *E*- and *H*-field distributions on a closed surface surrounding the scatterer [1,2]. In our approach, however, such mathematical manipulations become unnecessary, as we have direct access to the scatterer's internal oscillators, which oscillators produce the radiation not only in the far field but also throughout the entire space. The proof is completed by noting that the extinction cross-section and the forward scattering amplitude have identical expressions in terms of the scatterer's internal oscillations, thus obviating the need to self-consistently evaluate these oscillations in any detail at all.

Finally, it must be mentioned that the entire argument of this paper could have been made in terms of an effective charge distribution, $\rho_{\text{total}}(\boldsymbol{r}, t)$, and an effective current distribution, $\boldsymbol{J}_{\text{total}}(\boldsymbol{r}, t)$, induced within the material object by the incident plane-wave. Polarization and magnetization of the material, represented by their bound-charge and bound-current densities, would then have been subsumed within the corresponding total entities. In other words, in the approach presented in this paper, there is no need to expand the total charge and current distributions into various contributions from monopoles, dipoles, quadrupoles, etc. There are two reasons, however, why we chose to include in our analysis, in addition to free charges and free currents, the induced electric and magnetic dipoles via $\boldsymbol{P}(\boldsymbol{r}, t)$ and $\boldsymbol{M}(\boldsymbol{r}, t)$. First, Maxwell's *macroscopic* equations, which explicitly consider charge, current, polarization, and magnetization as sources of the electromagnetic field, do not include higher-order multipoles as distinct entities; so it seemed only natural to carry out the calculations with $\boldsymbol{P}$ and $\boldsymbol{M}$ as sources distinct from $\rho$ and $\boldsymbol{J}$—even though, in the end, the effects of $\boldsymbol{P}$ and $\boldsymbol{M}$ reduce to those of their corresponding bound charges and currents. Our second motivation for treating $\boldsymbol{P}$ and $\boldsymbol{M}$ explicitly, was that there exist two alternative treatments of the bound-current associated with magnetization, namely, as a bound electric current $\boldsymbol{J}_{\text{bound}}^{(\text{electric})}=\mu_o^{-1}\boldsymbol{\nabla}\times\boldsymbol{M}$, or as a bound magnetic current $\boldsymbol{J}_{\text{bound}}^{(\text{magnetic})}=\partial\boldsymbol{M}/\partial t$. In the first treatment, $\boldsymbol{J}_{\text{bound}}^{(\text{electric})}$ exchanges energy exclusively with the $E$-field, while in the second treatment $\boldsymbol{J}_{\text{bound}}^{(\text{magnetic})}$ exchanges energy only with the $H$-field. We opted here for the second treatment, because it leads to the definition $\boldsymbol{S}=\boldsymbol{E}\times\boldsymbol{H}$ of the Poynting vector, which many authors [1, 2, 11, 12] find more appealing than the first treatment, whose Poynting vector turns out to be $\boldsymbol{S}=\boldsymbol{E}\times\boldsymbol{B}/\mu_o$ [8]. Without engaging in a comparative analysis of the merits of these alternative Poynting vectors, we simply point out that the Optical Theorem remains unaffected by the method of treatment of magnetization within Maxwell's equations.